\documentclass[conference]{IEEEtran}
\IEEEoverridecommandlockouts

\usepackage{cite}
\usepackage{amsmath,amssymb,amsfonts}
\usepackage{algorithmic}
\usepackage{graphicx}
\usepackage{textcomp}
\usepackage[dvipsnames]{xcolor}

\usepackage{comment}
\usepackage[linesnumbered,ruled,vlined]{algorithm2e}
\usepackage{subcaption}

\def\BibTeX{{\rm B\kern-.05em{\sc i\kern-.025em b}\kern-.08em
    T\kern-.1667em\lower.7ex\hbox{E}\kern-.125emX}}
\begin{document}

\title{Characterizing the I/O Behavior of HPC Applications through Modeling and Simulation\\
\thanks{This work is has been partially funded by
the grant “Grants for the recruitment of predoctoral researchers-in-training for the year 2024.” with
reference PIPF-2024/COM-34495 funded by the Comunidad de Madrid and is part of the I+D+i project PID2022-
138050NB-I00 (New scalable I/O techniques for hybrid
HPC and data-intensive workloads - SCIOT), funded by
MICIU/AEI/10.13039/501100011033/ “FEDER A way to
do Europe”.}
}

\makeatletter
\newcommand{\IEEEauthorblocknewline}{%
  \end{@IEEEauthorhalign}
  \hfill\mbox{}\par
  \mbox{}\hfill\begin{@IEEEauthorhalign}
}
\makeatother

\author{
\IEEEauthorblockN{Njoud O. Almaaitah}
\IEEEauthorblockA{\textit{Department of Computer Science} \\
\textit{Mutah University}\\
Al-Karak, Jordan \\
njoudmaitah@mutah.edu.jo}
\and
\IEEEauthorblockN{David E. Singh}
\IEEEauthorblockA{\textit{Department of Computer Science and Engineering} \\
\textit{University Carlos III of Madrid}\\
Madrid, Spain \\
dexposit@inf.uc3m.es}
\and
\IEEEauthorblockN{Taylan \"{O}zden}
\IEEEauthorblockA{\textit{Department of Computer Science} \\
\textit{Technical University of Darmstadt }\\
Darmstadt, Hesse, Germany \\
taylan.oezden@tu-darmstadt.de}

\IEEEauthorblocknewline
\IEEEauthorblockN{Jesus Carretero}
\IEEEauthorblockA{\textit{Department of Computer Science and Engineering} \\
\textit{University Carlos III of Madrid}\\
Madrid, Spain \\
jcarrete@inf.uc3m.es}

\and
\IEEEauthorblockN{Raffaele Montella}
\IEEEauthorblockA{\textit{Department of Science and Technologies} \\
\textit{ University of Naples “Parthenope” }\\
Napoli, Italy \\
raffaele.montella@uniparthenope.it}

}

\maketitle

\begin{abstract}
Parallel applications process large amounts of data, leading to intensive parallel I/O operations. These operations can exhibit different levels of complexity, including, among others, multiple I/O access patterns, data staging, and contention risks. Therefore, in order to exploit high-performance computing (HPC) systems efficiently and optimize the I/O performance, it is crucial to consider the I/O behaviour of the HPC applications. 
In this work, we have developed a framework that reproduces the I/O access pattern of real applications in a simulated environment provided by ElastiSim, a batch-system simulator for rigid, malleable, and evolving workloads. 
The simulated applications are generated based on I/O traces captured from real applications provided by the HPC Input/Output (HPCIO) analysis repository. 
The HPCIO analysis database includes traces combined with information about real applications' performance across different parallel I/O libraries and layers of the I/O stack. We have conducted detailed case studies of real-world applications' traces to demonstrate how the proposed modeling framework can provide insights into the performance characteristics of I/O applications, including the I/O congestion analysis based on the application's I/O access pattern.
\end{abstract}


\section{Introduction}

HPC systems are designed to support the needs of compute/data-intensive applications. However, parallel applications often underperform on modern supercomputers due to several issues, such as load imbalance, resource contention, inefficient memory access, or network congestion. In particular, when we consider the I/O subsystem, applications are often restricted not only by the effectiveness of the I/O operations but also by potential contention hazards related to multiple applications performing simultaneous I/O accesses. This makes optimizing the application I/O a complex task, considering the existence of other running applications. 

Modeling I/O workloads is necessary for evaluating storage architecture and I/O software stack (e.g., burst buffer configurations or I/O scheduling algorithms). This helps to analyze the impact of workloads on storage system implementations and to identify performance bottlenecks~\cite{zhai2023performance,snyder2015techniques,behzad2019optimizing}. From the scheduling and task management side, the modeling of parallel applications improves scheduling decisions by helping the scheduler map the most appropriate nodes to a task \cite{carretero2020mapping}. However, the design of hardware and input parameters are among the factors that can affect the performance of parallel applications, making it challenging to create accurate performance models~\cite{snyder2015techniques,neuwirth2021parallel}.   
 
The proposed modeling framework uses the application profiles available in the HPCIO analysis repository~\cite{HPCIO}, a collaborative initiative to enhance knowledge and collaboration regarding the I/O aspects of HPC applications. The HPCIO analysis database includes data on application performance across different parallel I/O libraries and layers of the I/O stack. This work provides a methodology for porting the existing repository's traces to the ElastiSim~\cite{ozden2022elastisim} framework by creating a simulated application that reproduces the traces within the simulated platform. 
In this way, it is possible to simulate the application I/O behavior under different resource-level utilization. The main goal of this work is to support the development of more advanced I/O staging and scheduling techniques that are aware of the application I/O access patterns and leverage malleability both for improving the CPU and I/O execution times. As a proof-of-concept, we provide detailed case studies of real applications, and we demonstrate how the presented modeling framework is able to obtain insights into the platform I/O performance by considering the effect of contention when jobs with different I/O patterns are executed. 

To summarize, the major contributions of our work are:

\begin{itemize}
    \item A framework for reproducing I/O traces from the HPCIO portal by means of a novel methodology that generates simulated applications in ElastiSim with the same I/O pattern.

    \item An extensive validation of the \textit{ElastiSim} simulator using two representative parallel applications, \textit{Jacobi}~\cite{cascajo2024} and \textit{EpiGraph}~\cite{MERINO2023547}, executed and compared on two different real and simulated HPC platforms. 
    \item An evaluation of real use-case applications considering scenarios related to I/O congestion. 
\end{itemize}

The remainder of this paper is organized as follows: Section~\ref{sec:soa} provides an overview of the most related works. Section~\ref{sec:sp} presents the ElastiSim simulator as a simulation tool in this paper. In Section~\ref{sec:use}, real-world use-case applications that are addressed in this paper are briefly presented.  
 Section~\ref{sec:sf} presents the proposed modeling framework, including a description of the algorithms used for the modeling. Section~\ref{sec:ev} shows the performance metrics employed in the evaluation and the evaluation results of the framework. Finally, Section~\ref{sec:conc} summarizes the work and discusses future directions.
 
\section{Related work}\label{sec:soa}

Several techniques exist for modeling and predicting I/O application performance in HPC systems \cite{neuwirth2021parallel}. These techniques follow the approaches of statistics and analysis, predictive analytics, replay-based, or workload generation~\cite{neuwirth2021parallel}.  Statistics and analysis techniques help to extract meaningful patterns from numerical data by collecting, classifying, and representing it. The collected data usually include I/O traces and I/O characterization profiles~\cite{bez2023access}. Patel et al. in~\cite{patel2019revisiting} used this approach to perform a correlation analysis of the system to identify subtle relationships between different I/O activities and system components over time. MPItrace presented in~\cite{servat2010detailed} is a tracing tool that captures performance data of parallel applications. Paul et al.~\cite{paul2020understanding} collected statistics about the Lustre file system on compute nodes, metadata, and object storage file system servers.

In the predictive and analytics approaches, the direction is to identify performance issues in advance by analyzing historical data. For example, the research presented in~\cite{sun2020automated} relied on the fact that most high-performance computing (HPC) applications follow a pattern of initialization, repetitive processing, and termination phases to predict the applications' performance. 
Behzad et al. in~\cite{behzad2019optimizing} introduced an autotuning system designed mainly to hide the complexity of the I/O stack from the scientific application without compromising performance. 

Other studies utilize the replay-based approach, which involves analyzing the historical application traces to generate a copy of the workloads that can replay the I/O behavior of the original application~\cite{kim2020towards}. Luo et al. proposed a mathematical model and a replay engine for analyzing and extrapolating trace data and verifying the accuracy of the extrapolated trace file~\cite{luo2017scalaioextrap}. Snyder et al.~\cite{snyder2015techniques} proposed a technique to abstract HPC I/O workloads from different sources (e.g., Darshan or Recorder) to support I/O evaluation and analysis tools.

The applications simulated in this work are created based on I/O traces collected from real applications, which are available in the HPCIO analysis repository~\cite{HPCIO}. The goal of this work is to replicate the I/O access pattern of real applications in a simulated environment provided by ElastiSim. The proposed modeling framework provides insights into the performance characteristics of I/O applications, including the I/O congestion analysis based on the application's I/O access pattern.

\section{Simulation tool}\label{sec:sp}

As we aim to simulate HPC applications  
consisting of different tasks such as computation or I/O, and supporting performance models (i.e., mathematical functions) to describe their simulated workload, we chose ElastiSim as our simulation environment. ElastiSim is a batch-system simulator that focuses on rigid and elastic workloads such as malleable or evolving jobs. 

Although ElastiSim allows for the integration of custom scheduling algorithms, we leverage its features that focus on application simulation. ElastiSim is a discrete-event simulator written in C++, based on the widely-used platform simulation framework SimGrid~\cite{casanova2014simgrid}. Its workload model comprises jobs and so-called application models. While jobs specify attributes such as the job type (e.g., rigid or malleable) or the number of requested resources, application models specify the load introduced to the simulated platform. Application models in ElastiSim follow a strictly hierarchical structure consisting of phases and multiple tasks per phase, where each task represents a simulated activity, such as computation, communication, or an I/O operation. As ElastiSim allows us to simulate multiple applications concurrently on SimGrid platforms, comprising tasks of different types, it is a well-suited candidate to study applications that rely on shared resources, such as the parallel file system.

\section{Use-case applications} 
\label{sec:use}

To demonstrate the methods proposed in this paper, we have considered several use cases. The first two are Quantum Expresso distributions (Qe)~\cite{giannozzi2009quantum}. Expresso stands for opEn Source Package for Research in Electronic Structure, Simulation, and Optimization. The Qe is a collection of tools integrated for electronic-structure calculations and materials modeling. We have used two distributions, the Car-Parrinello (QeC), which uses density-functional theory for its calculations, and the PHonon (QePh) for vibrational properties, and the Density-Functional Perturbation Theory.

The third use case is Nek5000 (Nek) \cite{saha2021review}, an open-source solver created at the Argonne National Laboratory. One of the essential features of the solver is that it can solve equations without the need for intermediate communication steps. Nek's characteristics, such as the low memory required for simulations, make it highly efficient and scalable. 
WaComM++ (Water Community Model) is the fourth use case. It is an open-source software designed to help simulate and predict pollutant spills, transport, and dispersion in inshore and offshore environments. It is an essential component of a scientific workflow that enables the execution of numerical simulations, which provide high-resolution predictions of weather and marine conditions in the Bay of Naples. WaComM++ is an optimized version of the Lagrangian Assessment for Marine Pollution 3D (LAMP3D) model, featuring checkpoint and restart capabilities and shared memory parallelization~\cite{LUCCIO2017490}.

In addition, for validation purposes, we have used Jacobi and EpiGraph. Jacobi~\cite{cascajo2024} is the kernel of a numerical method for determining the solutions of a diagonally dominant system of linear equations. Each diagonal element is solved iteratively. EpiGraph~\cite{MERINO2023547} is an agent-based parallel simulator that performs realistic stochastic simulations of the propagation of the COVID-19 virus across wide geographic expanses. The considered application simulates two infection COVID-19 waves corresponding to Delta and Omicron variants in Spain.

\section{Modeling framework}\label{sec:sf}

The modeling framework, as illustrated in Figure~\ref{fig:overview}, consists of multiple phases. At the start, a web-based procedure collects the information that will subsequently used by the framework in the next phases. The available trace of each application (native I/O trace) consists mainly of data volumes read or written over a specific number of time intervals. Each interval (also called a time slot) represents a sample of the application execution (all intervals have the same duration) where performance and I/O metrics are collected. The execution environments of the selected applications (native platforms) were examined in terms of I/O bandwidth and computation power.
 
\begin{figure}[htbp] 
    \centering 
    \includegraphics[width=0.5\textwidth,angle=0, trim={1cm 2cm 1cm 3cm}, clip]{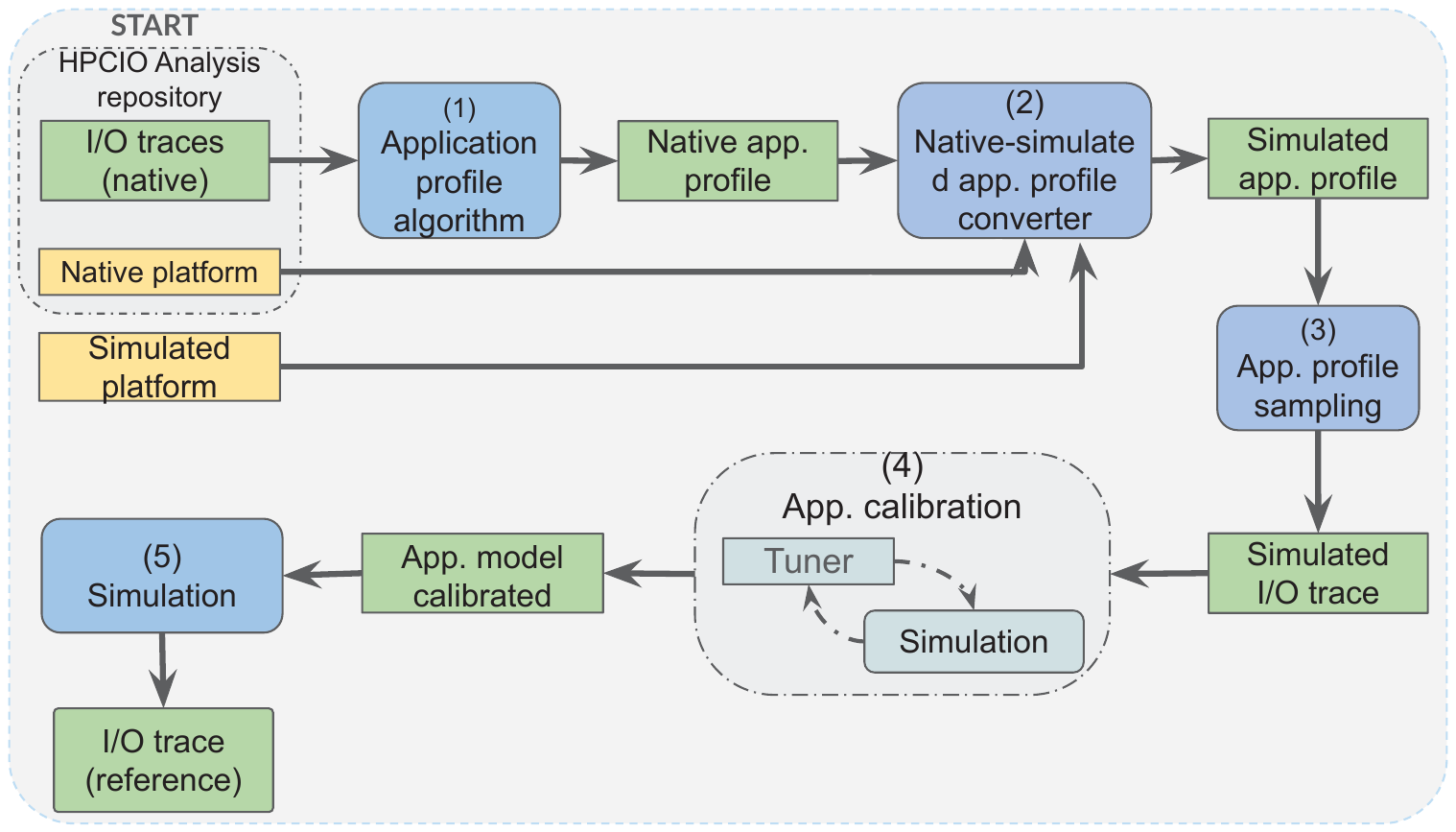}
    \caption{The modeling framework architecture overview.}
    \label{fig:overview}
\end{figure}

The application profile algorithm, step (1) in the overview,  analyzes the native application trace and estimates the duration of computation and I/O time within each interval. This work assumes that the CPU and I/O times are longer than the communication times, so the communication overheads are not considered in the current version of the framework (they will be addressed in future work). The result of this algorithm is the native application profile. 
It's important to note that the source (denoted as native platforms) and simulated platforms are different. 

Since the HPCIO repository includes multiple traces collected from different systems, our objective is to simulate them on a single platform, referred to as the reference platform. To achieve this, the original application behavior must be translated to match the architecture of this reference-simulated system.  
This approach permits us to compare multiple trace behaviors in a single execution environment. In order to achieve this objective, we use a native-to-simulated application profile converter, step (2) in the overview, which calculates the corresponding interval times over the simulated environment. 

The sampling algorithm in step (3) of the overview, Figure~\ref{fig:overview}, takes the simulated application profile as input and generates the application I/O trace that will be tested over the simulated platform, named sampled I/O trace. This step ensures accurate data distribution over the intervals. The I/O trace, until this phase, contains two tasks: the I/O task with sampled data amount and the computation task with a certain duration. The application calibration, step (4), receives the sampled I/O traces as input to perform tuning over the amount of computation load. The application model is calibrated once it behaves like native I/O traces, i.e., the application injects the same amount of I/O data volume at the same time as in the native trace. In step (5), the calibrated trace is tested through simulation to compare the simulated behavior with the native trace.

The next sections explain the work of each procedure and the models used in each step of the proposed framework.

\subsection{Trace importing and application profiles}

Our approach takes advantage of the data sources of Darshan summary~\cite{snyder2016darsh} provided by the HPCIO repository. The first is the records that provide I/O data volumes in specific time slots. The second is application statistics, such as average I/O bandwidth and execution time. The modeling framework imports the traces of the HPCIO repository; for example, Figure~\ref{fig:QE-PH} depicts the modeling phases of the Quantum Espresso PHonon (QePh) application. The native QePh trace from the HPCIO repository run in the JUWELS booster cluster is shown in Figure~\ref{fig:QE-PH}\subref{fig:qephNative}. The x-axis represents execution time in seconds, while the y-axis represents I/O data amounts in gigabytes.

\begin{figure}[t]
    \centering
    \begin{tabular}{@{}cc@{}}
       \begin{subfigure}{0.9\columnwidth}
            \centering
            \includegraphics[width=1\columnwidth]{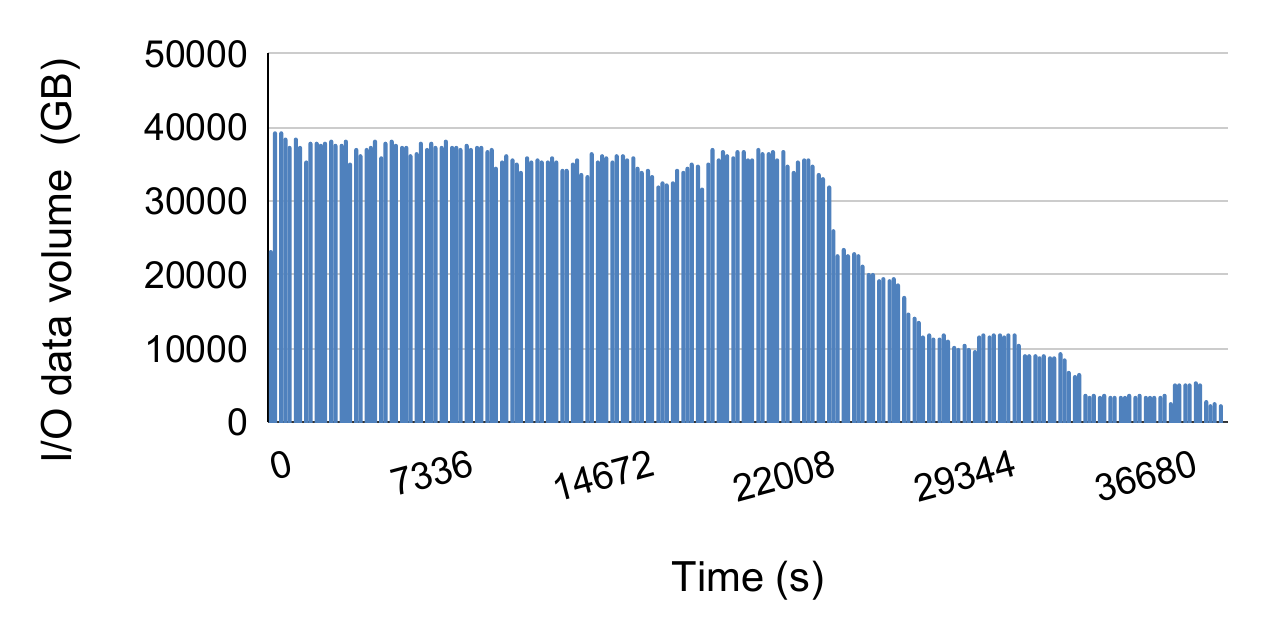}
            \caption{Native trace.}
            \label{fig:qephNative}
        \end{subfigure}
        \\        
        \begin{subfigure}{0.9\columnwidth}
            \centering
            \includegraphics[width=1\columnwidth]{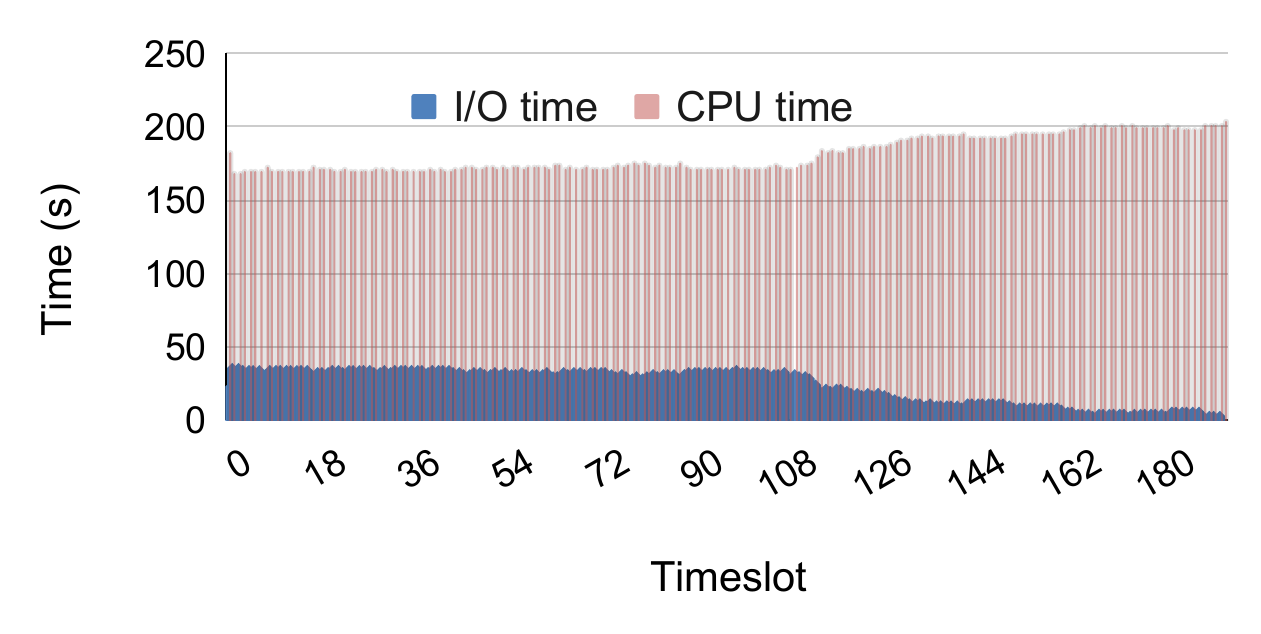}
            \caption{Native profile.}
            \label{fig:qephProfiled}
        \end{subfigure} \\
            \begin{subfigure}{0.9\columnwidth}
            \centering
            \includegraphics[width=1\columnwidth]{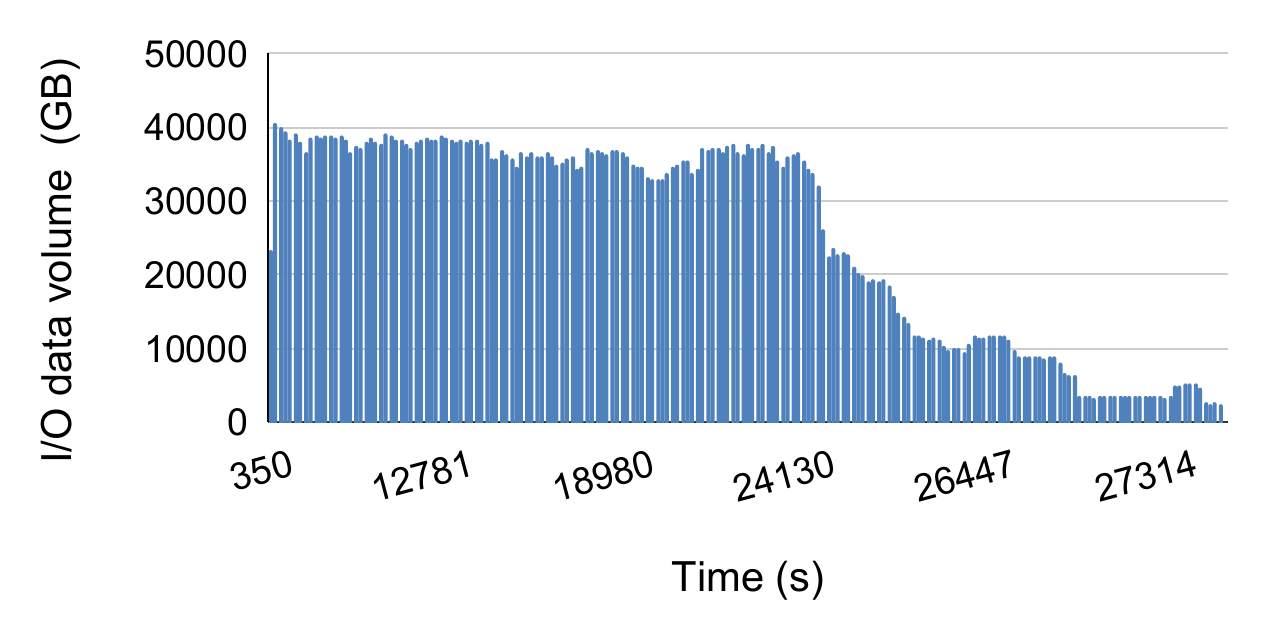}
            \caption{Simulated trace.}
            \label{fig:qephCali}
        \end{subfigure}
         \\
        \begin{subfigure}{0.9\columnwidth}
            \centering
            \includegraphics[width=1\columnwidth]{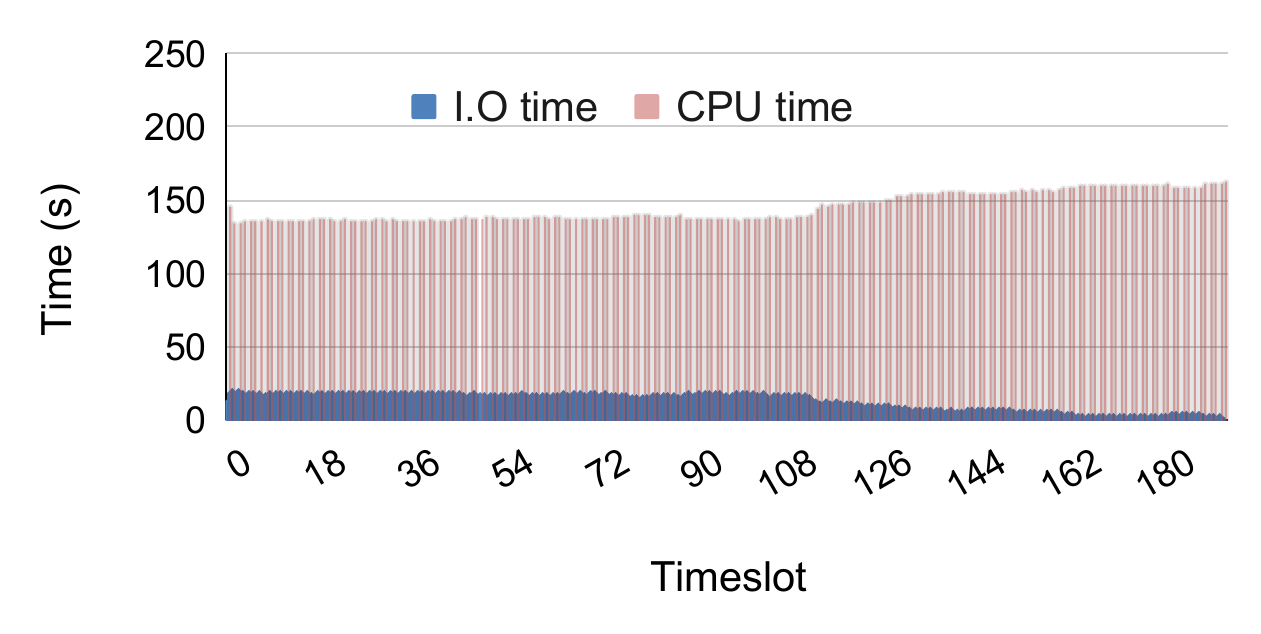}
            \caption{Simulated profile.}
            \label{fig:qephProfiled-sim}
        \end{subfigure}
       
    \end{tabular}
    \caption{QePh trace and profile comparison between: (a) and (b)  native execution on the JUWELS booster cluster; (c) and (d) simulated execution on the reference platform.}
    \label{fig:QE-PH}
\end{figure}

The HPCIO repository provides detailed information about I/O traces. This information includes timestamps indicating when each read/write operation occurs. Incremental timers measure I/O time, while histograms visualize I/O access patterns. However, the information related to I/O application performance, such as average throughput, is used for native and simulated application profiles in our framework. The application profile algorithm computes the interval length $T_{i}$ of the collected trace and the duration of the CPU and I/O phases of each interval. The I/O phase duration of the interval $i$,  \( T^{\text{IO}}_{i}\) is computed by dividing the interval I/O data over the native I/O bandwidth (provided by the Darshan profile at the HPCIO repository). The CPU phase duration \( T^{\text{CPU}}_{i} \) is the remainder time of the interval. Figure~\ref{fig:QE-PH}\subref{fig:qephProfiled} shows the profiling of the native trace with QePh application.

When we port the trace from the native platform into the simulated one, the execution times of the CPU and I/O interval may differ because of differences in the CPU and I/O performances between both platforms (the JUWELS booster cluster and the reference one). The application profile uses the ratios of  CPU processing power and I/O bandwidths between native and simulated platforms ($R_{cpu}$ and $R_{io}$, respectively) to estimate the execution time of each CPU and I/O interval on the simulated platform.  This produces a profile for the simulated application. For example, as illustrated in Figure~\ref{fig:QE-PH}\subref{fig:qephProfiled-sim}, when we run an application on a faster platform ($R_{cpu}$ = 0.8 and $R_{io}$ = 0.5), the CPU and I/O times are reduced, in such a case.

\subsection{Application profile sampling and model calibration}

The next step is to generate a simulated I/O trace based on the application profile. The sampling algorithm processes the application profile data at specified intervals (denoted as the sampling window). One sampling window could include one or more intervals. The sampling procedure takes three inputs: the amount of data to be sampled, the sampling window's start time, and the end time of the sampling window. It iterates through each interval entry, where the sample size is determined based on the execution time of the profiled application and the number of intervals. For example, as illustrated in Figure~\ref{fig:exsample}, if an interval falls entirely within the current sample window (intervals \textbf{A} and \textbf{B} in the figure), its data is added to the sample data. If the interval spans multiple sample windows (interval \textbf{C} in the figure), the sampling procedure distributes data proportionally between the overlapped sample windows. In this case, it calculates the fractions of the interval that fall within each window ($s_{1}$ and $s_{2}$), computes the corresponding data amounts ($v_{1}$ and $v_{2}$), and updates the sample data accordingly for the current and next sample windows. As shown in Figure~\ref{fig:exsample}, the sample windows 1, 2, and 3 will have I/O data volumes of 107, 126, and 57 gigabytes, respectively. Note that we assume a homogeneous distribution of I/O data within each interval. The output of the sampling procedure is the simulated I/O trace.


\begin{figure}[t]
            \centering  
            \includegraphics[clip=true, trim={1cm 4.7cm 0cm 0cm},width=0.9\columnwidth]{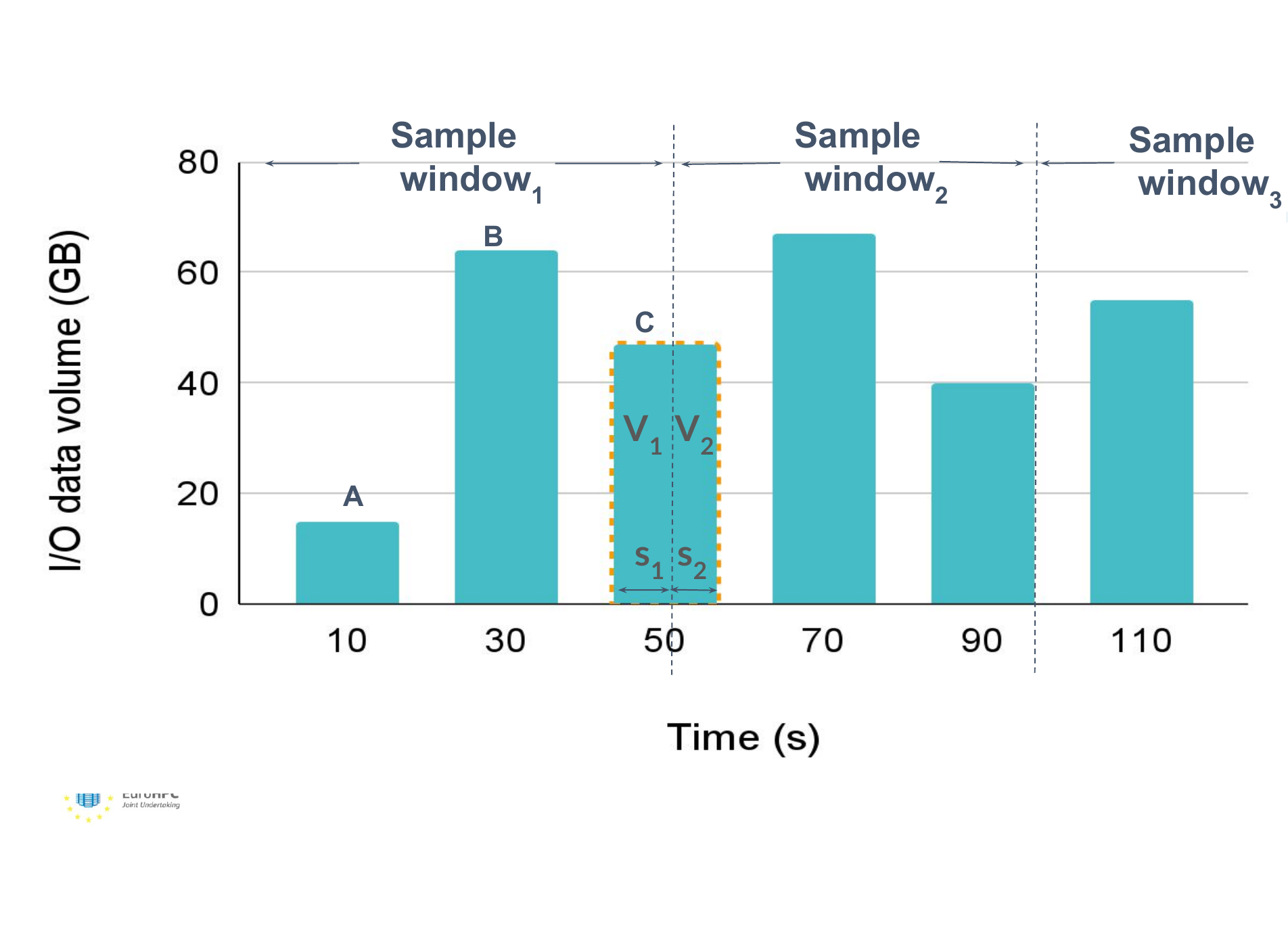}
             \caption{Example where the interval simulated time is larger than the original. The application trace sample window is 52 seconds, and the original interval is 20 seconds.}
            \label{fig:exsample}
\end{figure}\hspace{0.5cm}

The application model calibration, Algorithm 1, 
adjusts the computation phase of the simulated application to produce the same I/O trace as the one received as input. The trace represents a sequence of intervals that alternate CPU and I/O operations. Initially, the algorithm sets up a simulated application that consists of as many phases as I/O trace intervals. The I/O operation of a certain phase, in any given trace, processes the same amount of data as one of the related intervals in the simulated trace. Regarding the CPU operation in each phase, the application model starts with an initial amount of computation, expressed as a number of Flops. Then, each CPU phase's duration is adjusted in an iterative manner to meet the same duration as the one in the simulated I/O trace. 
If the observed task duration exceeds the one in the related interval, then the Flops for that phase are reduced. Otherwise, the Flops are increased when the duration is smaller than the one in the interval. These changes are only applied in each application phase if the observed deviation is above a certain threshold. 
This procedure is carried out iteratively by running multiple simulations and refining the application model. For each run, the cumulative deviation of each execution is computed. The error estimates how the overall internal length differs from the interval length in the sampled application. The process is considered completed when the error is under a given threshold. The result of this phase is an application model that can be simulated in ElastiSim. In the QePh use case, the I/O trace produced by the calibrated application is represented by Figure~\ref{fig:QE-PH}\subref{fig:qephCali}. Note that the resulting trace is similar to the native trace but with a shorter execution time as a consequence of a faster simulated platform ($R_{cpu}$ = 0.8 and $R_{io}$ = 0.5). 

\begin{algorithm}[t]
\caption{Application Model Calibration}
\SetAlgoLined
\SetKwInOut{Input}{Input}
\SetKwInOut{Output}{Output}

\Input{task\_times, application\_model, threshold}
\Output{Updated application model}

\texttt{elastiSim.run()} \;
GET total\_phase\_times, execution\_time \;
size $\leftarrow$ \texttt{|phase\_durations|}, \ Ts = \texttt{trace\_time}/size \;
err $\leftarrow$ \texttt{Calculate(errors)} \;

\While{err $>$ threshold}{
    GET I/O\_times, phase\_durations \;
    SET initial \texttt{Flops}, $t{=}0.02$, $dT{=}0.05$ \;
    \For{i $\leftarrow$ 1 \KwTo size}{
        $t_i$ $\leftarrow$ phase\_duration$_i$ \;
        \If{$t_i > Ts(1{+}t)$}{\texttt{flops[i]} $\leftarrow$ \texttt{flops[i]}(1$-$$dT$)} \;
        \If{$t_i < Ts(1{-}t)$}{\texttt{flops[i]} $\leftarrow$ \texttt{flops[i]}(1$+$$dT$)} \;
    }
    \texttt{update\_app\_model(application\_model)} \;
    \texttt{elastiSim.run()} \;
    err $\leftarrow$ \texttt{Calculate(errors)} \;
}
\label{algo:calibrate}
\end{algorithm}

\section{Evaluation}\label{sec:ev}

In this section, we describe the reference simulated platform used for the experimental study of the applications. Beyond analyzing the modeled application behavior under interference in the provided scenarios, we also validated the simulation by comparing the simulated execution times with the real execution times of benchmark applications (\textit{Jacobi} and \textit{EpiGraph}). 

\subsection{System platform characterization}

When simulating HPC systems, it is necessary to model the hardware components that affect application performance and I/O behavior. The considered reference platform includes 4300 nodes, and each node contains 48 cores. Assuming that each processor is of the Intel Xeon Platinum 8160 24C type at 2.1 GHz, the node peak performance power is 3.2 Tflops. The parallel file system (PFS) is modeled as one shared storage tier accessible by all compute nodes with the type MareNostrum 4 GPFS (4.2.2.0) and GekkoFS (0.9.1). The platform has maximum transfer rates per node of 8 GB/s and 5 GB/s dedicated to reading and writing operations, respectively. The bandwidth configurations of the simulated platform rely on the findings that are presented in~\cite{garcia2023new} for MareNostrum 4. The simulated network connection is connected with a maximum transfer rate of 12.5 GB/s. The parallel file system bandwidth is 250 GB/s. 

\subsection{Application modelling}

As mentioned before, we have considered several use cases: The Quantum Expresso-PHonon (QePh), the Quantum Expresso Car-Parrinello (QeC), Nek5000 (Nek), and WaComM++.

Figure~\ref{fig:appplots} illustrates the results of the simulation of the four applications. The z-axis represents the I/O data volume produced by the application for each time interval. The y-axis corresponds to the number of nodes used in each simulation. For instance, a y-axis value of 32 is related to a simulation that generates an I/O trace using 32 nodes.
The x-axis corresponds to the number of intervals in the I/O trace. In each trace, the number of intervals remains the same despite the number of nodes that have been used. This is a way of normalizing the visualization of the I/O traces for a different number of nodes. In other words, all executions represented in the y-axis produce traces that have the same length, even though the execution time is reduced when increasing the number of nodes.

 \begin{figure}[!t]
 \vspace{-0.5cm}
    \centering
    \begin{tabular}{@{}cc@{}}
           \begin{subfigure}{0.45\columnwidth}
            \centering
            \includegraphics[width=1\columnwidth]{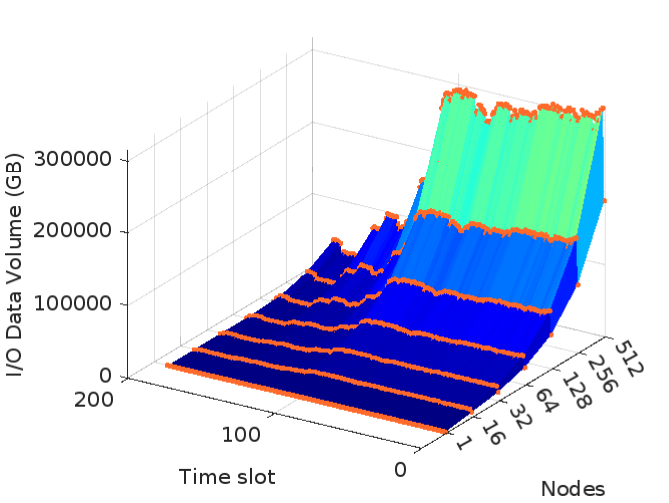}
            \caption{QePh}
            \label{fig:qecp-plot}
        \end{subfigure} &
       \begin{subfigure}{0.45\columnwidth}
            \centering
            \includegraphics[width=1\columnwidth]{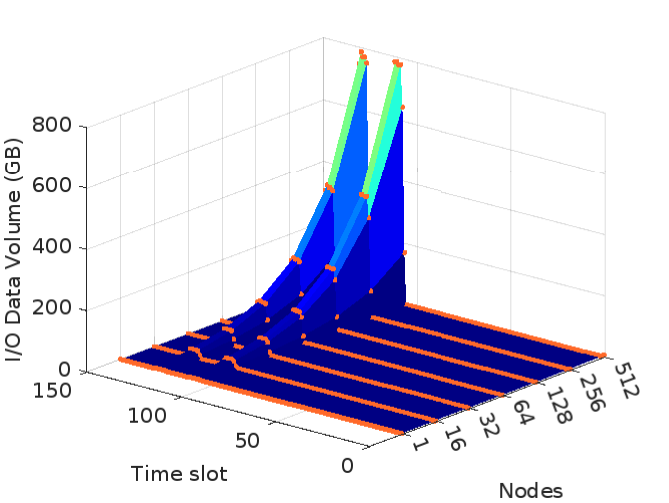}
            \caption{QeC}
            \label{fig:qeCp-plot}
        \end{subfigure}
        \\        
       \begin{subfigure}{0.45\columnwidth}
            \centering
            \includegraphics[width=1\columnwidth]{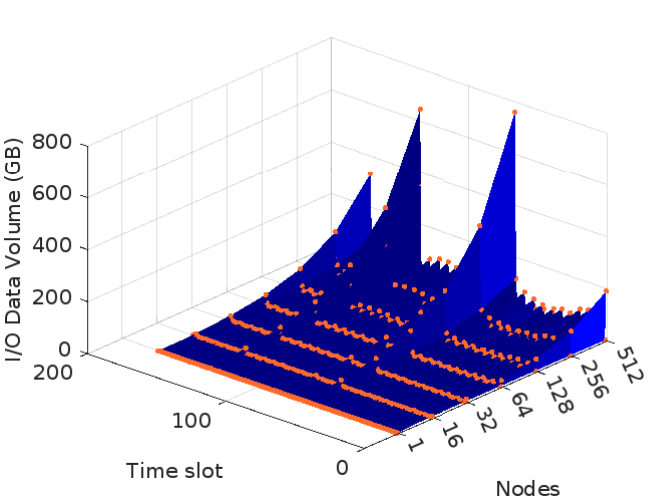}
            \caption{Nek}
            \label{fig:nek-plot}
        \end{subfigure} &
        \vspace{1cm}
        \begin{subfigure}{0.45\columnwidth}
            \centering
            \includegraphics[width=1\columnwidth]{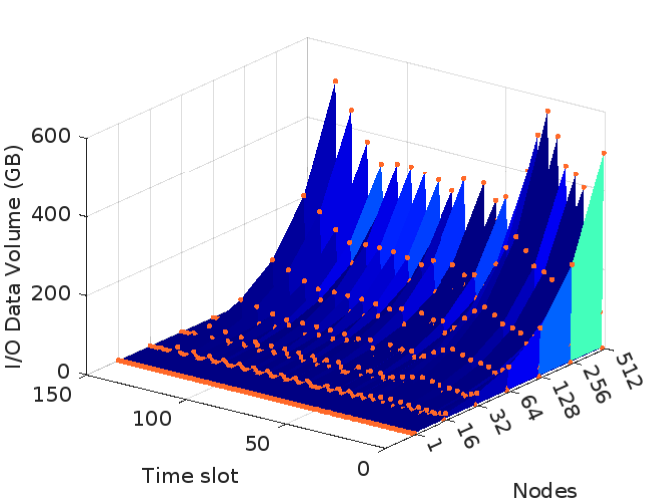}
            \caption{ WaComM++}
            \label{fig:WaCommplot}
        \end{subfigure}
    \end{tabular}
    \vspace{-0.5cm}
    \caption{I/O data volume for the four use-cases executed with different numbers of nodes and uniform data distribution pattern.  }
    
    \label{fig:appplots}
\end{figure}

\subsection{Framework and ElastiSim validation}

To validate the modeling capabilities of the proposed framework and the ElastiSim simulator, we evaluated how an application model produced for a certain platform (HPC4AI cluster~\cite{HPC4AI}, University of Turin) has been used to reproduce the application behavior for a different machine, the Center for Scientific Computing Cluster (C3) \cite{uc3m_c3} at University Carlos III de Madrid (UC3M). For this purpose, 
we captured execution traces from two distinct applications, \textit{Jacobi} and \textit{EpiGraph},
running on HPC4AI cluster. Based on those traces, the application model was subsequently obtained for each one of them. Then, the models were executed on the simulated destination C3 platform, and the simulated application behaviours were compared with the real ones on the C3.

For the \textit{Jacobi} and \textit{EpiGraph} applications, we configured them to run on 8 and 1 compute nodes in the HPC4AI cluster, respectively. Each compute node runs 36 processes. In both cases, we used the Darshan I/O tracing tool to capture detailed execution traces, including I/O events. The captured traces were analyzed using our framework, as described in Section~\ref{sec:sf}, where the framework produces traces compatible with the ElastiSim simulator. Based on these traces, an application performance model was generated for both applications. Subsequently, we configured the simulated platform in the simulator with the hardware and network parameters shown in Table~\ref{tab:Valplatform-features}. Using this configuration, we simulated both models for the C3 cluster, generating the new execution traces for this platform. 

\begin{table}[t]
\centering
\caption{Main features of the reference simulated platform for the validation}
\label{tab:Valplatform-features}
\resizebox{\columnwidth}{!}{%
\begin{tabular}{l c l}
\hline
Feature & Value & Notes \\
\hline
Number of nodes & 100 & - \\
Cores per node & 48 & - \\
Peak node performance & 1 TFLOPS & Node-level computation \\
Node bandwidth & 8 GB/s & Maximum transfer rate per node \\
Interconnect bandwidth & 100 Gbps & -  \\
Parallel File System & 256 GB/s & Shared PFS accessible by all nodes \\
\hline
\end{tabular}%
}
\end{table}

In order to validate our proposal, we compared key behavioral metrics such as execution time 
from the simulated traces against the real ones in C3 cluster at UC3M. Figures~\ref{fig:frameElsVal}(a) and~\ref{fig:frameElsVal}(b) illustrate the degree of similarity between the real and simulated executions for both applications. Each bar, in both figures, represents each application I/O phase. The x-axis and y-axis values corresponds to the I/O timestamp and accuracy of the execution time of each phase, respectively. Note that \textit{Jacobi} has a periodic I/O pattern, and in this case both bars (simulated and real I/O phases) almost completely overlap. This is the reason why in Figures~\ref{fig:frameElsVal}(a) only the simulated bars are shown (the real ones are hidden just behind them).

\textit{EpiGraph} has a non-periodic pattern related to the simulation of COVID-19 waves. For this application, when the Delta and infection waves are simulated, the number of infections sharply increases, incrementing the computation and communication operations performed by the simulator and spacing the I/O operations that are periodically done every certain number of iterations. That is to say, the I/O is periodically performed, but the iteration time changes according the simulation conditions (number of infections). Before 1100 secs  in Figure  ~\ref{fig:frameElsVal}(b), the COVID-19 incidence is low, so the I/O operation are performed frequently, because of a low CPU and communication intensity. Omicron and Delta waves are simulated around execution time 1100 secs, producing a significant gap between the I/O checkpointing between 1100 sec. and 1950 secs. Finally, the disease incidence remains high, producing for the rest of the simulation (times larger than 2000 secs) larger gaps than at the beginning of the simulation. Note that despite this complex I/O pattern, the performance model generated in HPC4AI cluster is able to reproduce the I/O pattern for the C3  cluster. 

The results indicate that the proposed framework, together with ElastiSim, successfully reproduces the performance behavior of both applications with accuracy reaches up to 98.8\% and 98.1\%.

\begin{figure}[t]
\centering
    \begin{tabular}{@{}cc@{}}   
    \begin{subfigure}{0.9\columnwidth}     
            \centering
            \includegraphics[width=1\columnwidth]{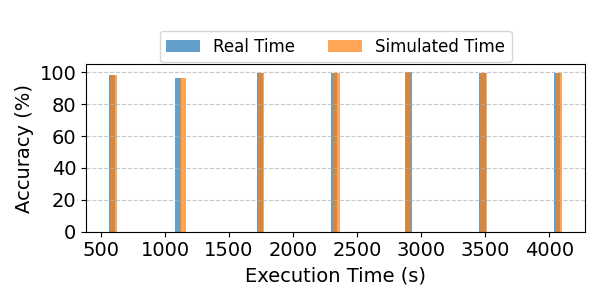}
            \caption{Jacobi}
            \label{fig:jac-exeTime}
        \end{subfigure}
        \\
        \begin{subfigure}{0.9\columnwidth}
            \centering           \includegraphics[width=1\columnwidth]{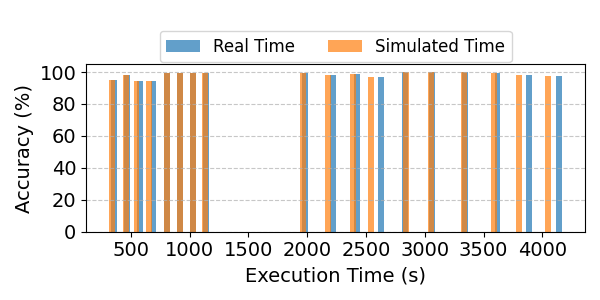}
            \caption{EpiGraph}
            \label{fig:epi-exeTime}
        \end{subfigure} 
        \end{tabular}
    \caption{Simulated and real execution traces for (a) Jacobi, and (b) EpiGraph on the C3 cluster at UC3M.}
    \label{fig:frameElsVal}  
     \end{figure}

\subsection{Performance evaluation}

When applications compete for system resources, it can cause interference, resulting in reduced performance and poor system utilization.  
First, the application's I/O throughput was studied by running experiments with multiple parallel jobs. For the sake of simplicity, each experiment has jobs of the same type, and all the jobs allocate the same number of nodes\footnote{The simulation granularity of ElastiSim is a node (consisting of 48 cores), so the results include this number as a scalability metric.}. The jobs have different submit times, which permits the overlap of the computation and I/O phases between the jobs. There is no resource over-subscription, and the simulated platform has more resources (CPU cores) than the number of processes used by all running applications. 

Figure~\ref{fig:inter} shows how the I/O throughput of the application types QePh and WaComM++ is affected when an increasing number of parallel jobs is executed. The figure displays the I/O application throughput of only one of the running applications. It is important to highlight that in order to compare different executions, the x-axis represents the normalized execution time. In Figure~\ref{fig:inter}\subref{fig:qeph-inter}, the file system bandwidth of 250 GB/s is reached when two jobs are simultaneously executed. As the number of jobs increases, the application throughput decreases because they compete for the available I/O bandwidth, producing contention in the parallel file system. We denote it as I/O interference. The nature of QePh indicates that the amount of I/O data volume near 80\% of execution time is smaller than the one before (see I/O pattern in Figure~\ref{fig:QE-PH}). Accordingly, we can observe that the I/O throughput between 80\% and 100\% of the run time does not degrade to the same extent as in the previous period because there is less contention during the last part of the execution. Figure~\ref{fig:inter}\subref{fig:wac-inter} shows a similar analysis for WaComM++. The I/O throughput in the case of WaComM++ is affected only when a much bigger number of jobs are executed simultaneously. This is because the WaComM++ I/O access pattern involves much less data traffic than the previous application. Note that by means of the framework presented in this work, it is possible to detect contention both related to the number of existing jobs and the related application I/O patterns that are being used.

\begin{figure}[t]
    \centering
    \begin{tabular}{@{}c@{}c@{}}  
        \begin{subfigure}{0.50\columnwidth}
            \centering           \includegraphics[width=1\columnwidth]{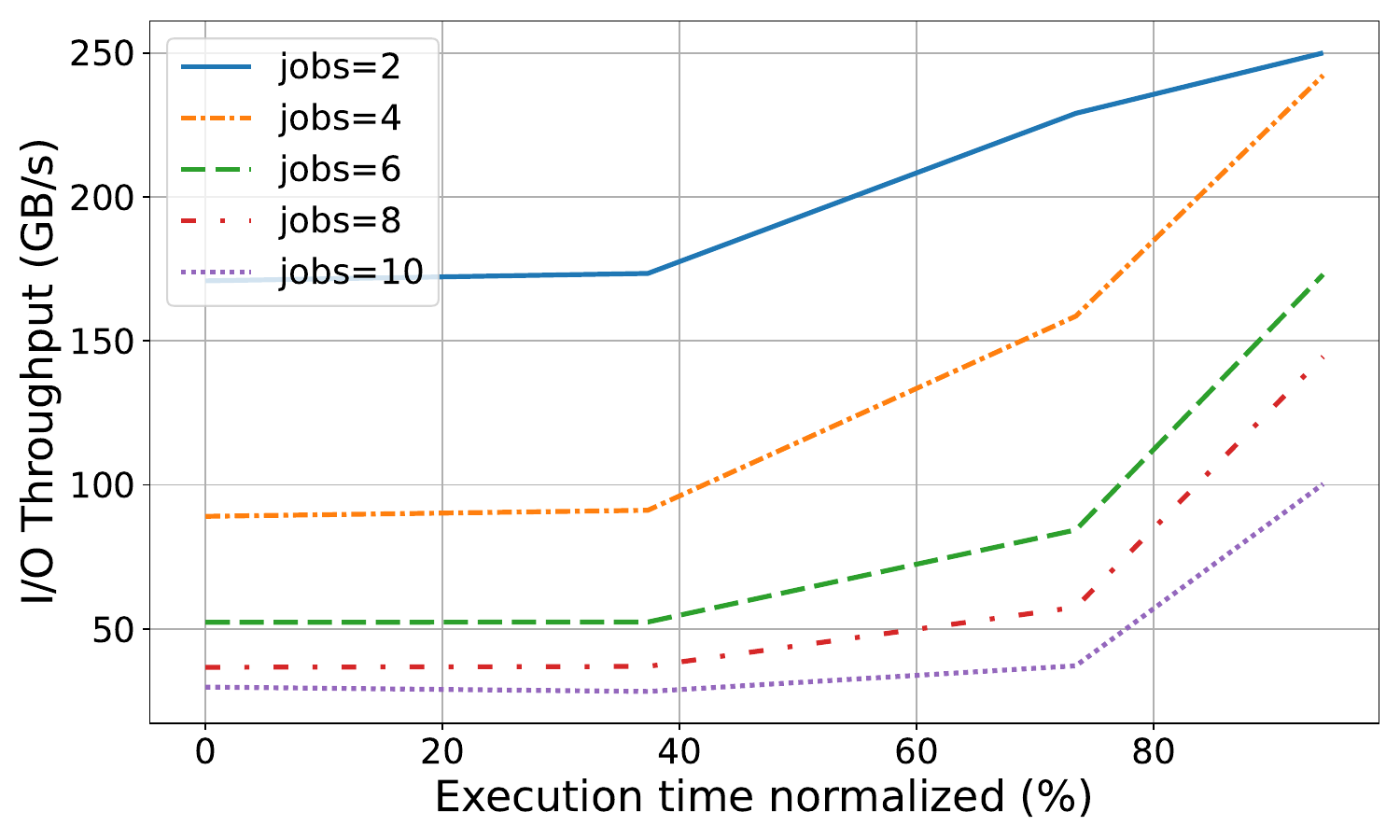}
            \caption{QePh, each job is located over 128 nodes.}
            \label{fig:qeph-inter}
        \end{subfigure} 
        &
        \begin{subfigure}{0.50\columnwidth}     
            \centering
            \includegraphics[width=1\columnwidth]{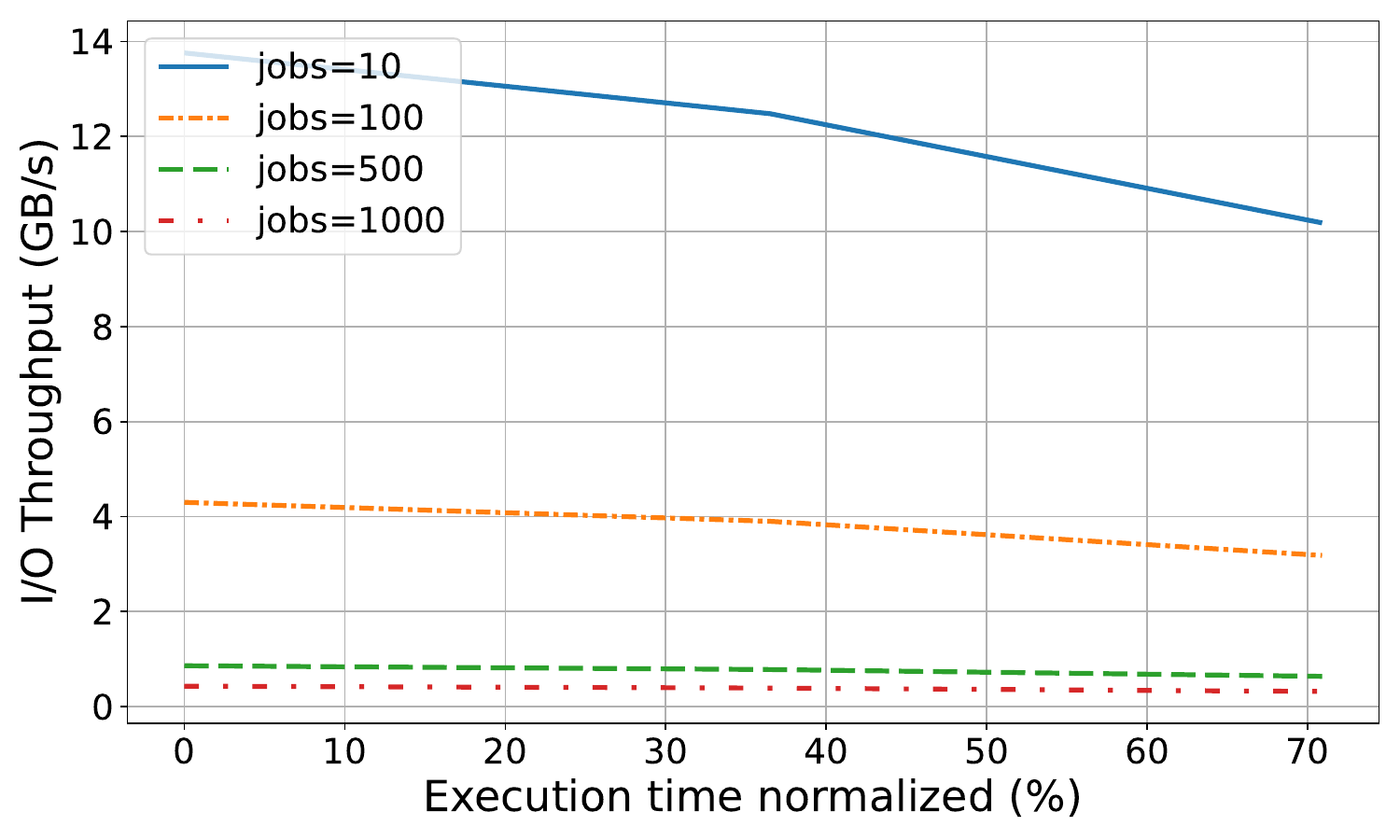}
            \caption{WaComM++, each job is located over two nodes.}
            \label{fig:wac-inter}
        \end{subfigure}
        \end{tabular}
    \caption{I/O interference analysis of (a) QePh and (b) WaComM++ for an increasing number of simultaneously running applications. The displayed I/O 
    throughput corresponds to only one of the running applications. }
    \label{fig:inter}   

\end{figure}


The second scenario, depicted in Figure~\ref{fig:qeph-inter-patterns}, explores another dimension of the problem: the I/O interference analysis when the job size increases, keeping the number of running applications fixed. In this case, we simulate the concurrent execution of four jobs of QePh type under two different data distributions: uniform and distributed. Each series corresponds to the execution of the application with a certain number of nodes. For instance, n=64 means the four applications are executed in 64 nodes each (256 nodes in total). Figure~\ref{fig:qeph-inter-patterns}\subref{fig:qeph-inter-io-uni} shows how the I/O throughput of one of the QePh jobs when executed with a uniform distribution pattern. As the data is replicated, there is more pressure on the I/O subsystem when the number of nodes increases, dramatically decreasing the application I/O throughput when the number of nodes increases.  Even after 80\% of execution time (when the I/O volume is reduced). 
When we consider the distributed pattern (Figure~\ref{fig:qeph-inter-patterns}\subref{fig:qeph-inter-io-d}), there is also a degradation in the I/O throughput when the number of nodes increases. This is because the aggregated application I/O throughput is bigger than the parallel filesystem throughput. However, the contention is 
smaller in the last 20\% of the application execution time because the data is now distributed over the nodes, so there is less pressure on the filesystem. 

\begin{figure}[t]
    \centering
    \begin{tabular}{@{}c@{}c@{}}  
         \begin{subfigure}{0.50\columnwidth}     
            \centering
            \includegraphics[width=1\columnwidth]{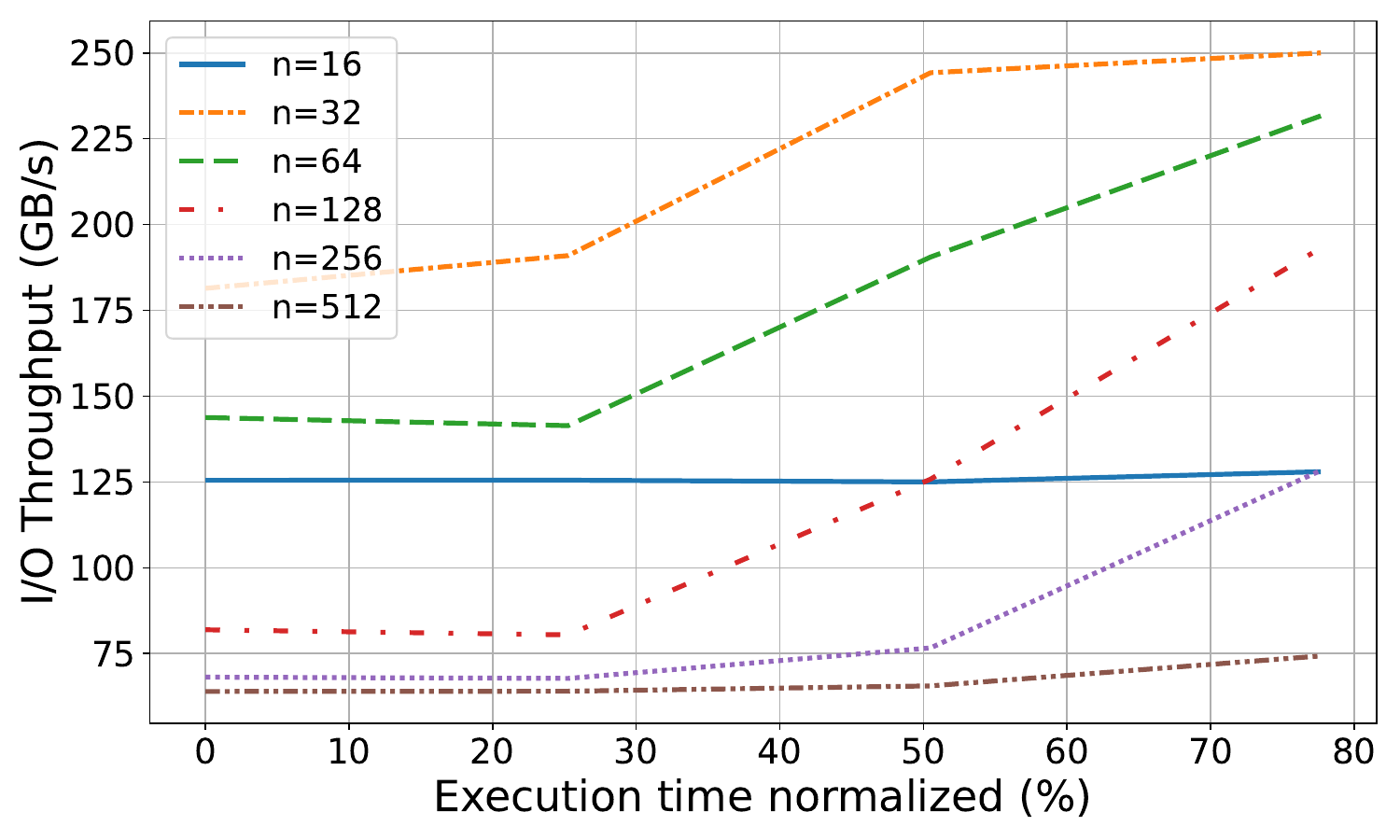}

            \caption{Uniform}
            \label{fig:qeph-inter-io-uni}
        \end{subfigure} 
        &
        \begin{subfigure}{0.50\columnwidth}     
            \centering
            \includegraphics[width=1\columnwidth]{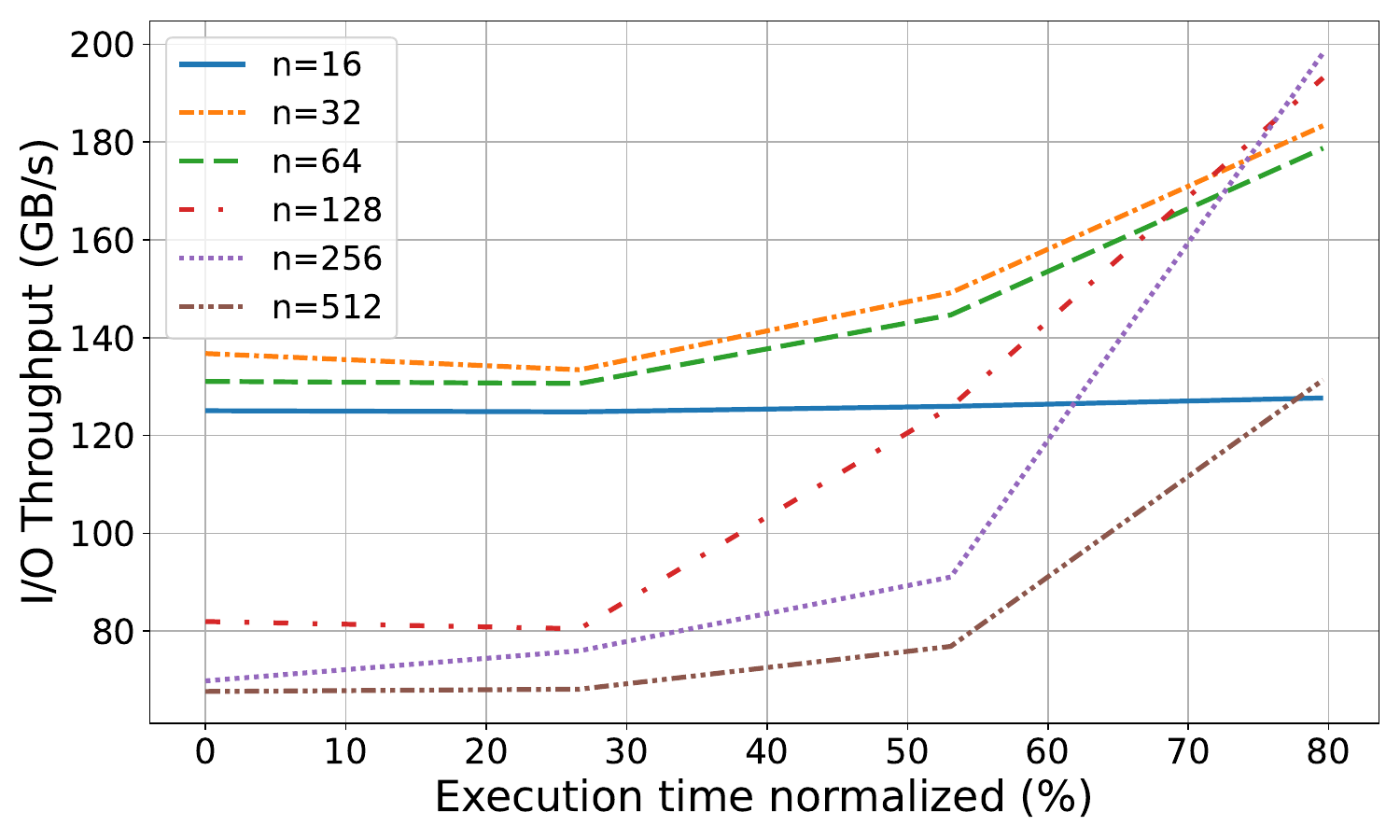}
            \caption{Distributed}
            \label{fig:qeph-inter-io-d}
        \end{subfigure} 
         \end{tabular}
         \vspace{-0.3cm}
 \caption{The I/O throughput when four applications are executed simultaneously and each application runs between 16 and 512 nodes 
 with (a) Uniform pattern and (b) Distributed pattern. The displayed I/O 
    throughput corresponds to only one of the running applications.}
    \label{fig:qeph-inter-patterns}
\end{figure}


In the following experiments, we extend our analysis to workloads composed of different use cases. We simulate the execution of different numbers of simultaneous jobs, each running on 64 nodes with a distributed data pattern. Given that in this pattern, the data volume does not increase with the number of nodes, the QePh and Nek's applications are the most I/O intensive jobs due to the considerable I/O data volume that they produce as a total and data distribution over intervals. For instance, the total I/O volume for QePh is 4812242 gigabytes, compared to WaComM++ and QeC (53 and 1117 gigabytes, respectively). In our experiments, we have considered three different workload profiles with a combined percentage of QePh and Nek of 10\%, 25\%, and 50\% in each. This means that half of the running jobs in the 50\% workload profile will be either QePh or Nek. Figure~\ref{fig:wAvg-degredation} 
shows the I/O degradation percentage for each one of these profiles and an increasing number of jobs per profile, all of which are simultaneously executed. The I/O degradation percentage is computed as the decrease in I/O throughput compared to the application executed exclusively. For instance, if the exclusive I/O throughput is 100 GB/s and it is decreased to 50 GB/s when executed with other jobs (as part of a workload), then the I/O degradation percentage will be 50\%. This figure shows the average I/O degradation among all the executed applications. We can observe that when QePh and Nek represent 10\% of the workload, the degradation starts with 32 jobs running concurrently, and the degradation percentage, in this case, does not exceed 20\%. When the percentage of I/O intensive jobs increases, 25\% and 50\% workload profiles, the I/O degradation occurs with a smaller number of jobs and has a larger extent. This reflects how different I/O access patterns impact the overall file system performance.

\begin{figure}[t]
    \centering
    \includegraphics[width=0.7\columnwidth]{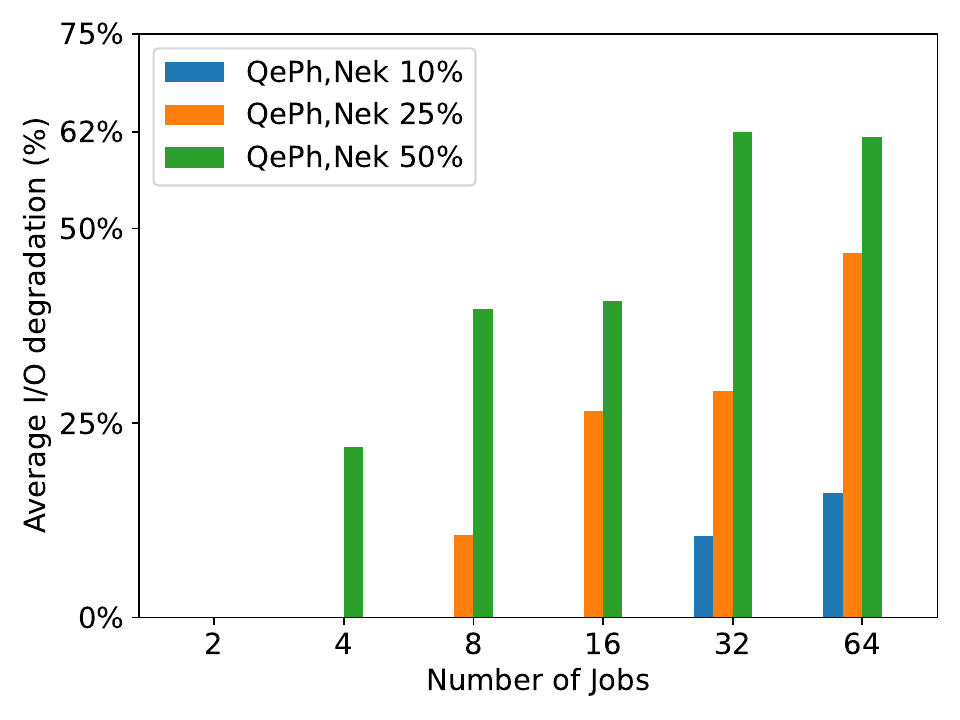}
    \caption{Average bandwidth degradation of workloads under a distributed data pattern. 
    }
    \label{fig:wAvg-degredation}
\end{figure}

\section{Conclusion}\label{sec:conc}

This paper proposes a framework to create applications in the ElastiSim simulator that reproduce real-world I/O traces provided by the HPCIO repository. 
In this work, we provide an exhaustive evaluation considering real-world application use cases. Our results show that the proposed framework provides feasibility and accuracy in reproducing the native application in a simulation environment and permits the obtaining of fine-grained insights about the platform performance, from the I/O perspective. For future work, there are two main research directions. The first one is simulating the malleable jobs and developing malleable I/O scheduling techniques that are aware of the running application's scalability and I/O access pattern. The second one is to extend the simulated platform with burst buffers in the I/O system and to use more refined performance models, obtained from tools such as Extra-P, to analyze the native I/O application scalability.


\bibliography{bibo}

\end{document}